\documentclass[12pt]{iopart}
\usepackage[dvips]{epsfig}

\newcommand{\figwidth}{0.8\textwidth}
\newcommand{\bacusio}{BaCu$_2$Si$_2$O$_7$}
\newcommand{\myvec}[1]{\mathbf{#1}}
\newcommand{\vectorprod}[2]{\bigl[\mathbf{#1}\times\mathbf{#2}\bigr]}
\newcommand{\scalarprod}[2]{\bigl(\mathbf{#1}\cdot\mathbf{#2}\bigr)}

\begin{document}
\title{Multiple spin-flop phase diagram of \bacusio.}
\author{V.N.Glazkov$^{1,2}$, G.Dhalenne$^3$ and
A.Revcolevschi $^3$, A.Zheludev$^1$}
\address{$^1$ Laboratorium f\"{u}r Festk\"{o}rperphysik, ETH Zurich, 8093 Z\"{u}rich, Schweiz}
\address{$^2$ P.~L.~Kapitza Institute for Physical Problems RAS, 119334  Moscow, Russia}
\address{$^3$ Laboratoire de Physico-Chimie de l'Etat Solide, Universit\'{e} Paris-Sud, 91405 Orsay cedex, France}
\ead{glazkov@kapitza.ras.ru}

\begin{abstract}
The quasi one-dimensional compound \bacusio{} demonstrates
numerous spin-reorientation transitions both for a magnetic field
applied along the easy axis of magnetization, and for a magnetic
field applied perpendicular to it. The magnetic phase diagram for
 all three principal orientations is obtained by magnetization
and specific heat measurements. Values of all critical fields and
low-temperature values of magnetization jumps are determined for
all transitions.
\end{abstract}


\submitto{\JPCM}

\section{Introduction.}

A spin-flop transition is a well-known characteristic feature of
easy-axis antiferromagnets. It is caused by a competition between
the anisotropy energy, which is minimized by aligning the order
parameter along the easy axis, and the Zeeman energy, which is
minimized for an antiferromagnet by aligning the order parameter
perpendicularly to the field. Thus, at a certain field value, the
order parameter re-orients itself, going away from the easy axis.
For a collinear antiferromagnet only one transition of such sort
is usually observed when the magnetic field is applied along the
easy axis.

The quasi-one-dimensional compound \bacusio{} has attracted much
interest due to its "excessive" spin-flop transitions. The main
exchange integral in this compound is J=24.1meV, while inter-chain
interactions are a factor of 100 smaller
\cite{kenzelman-prb-2001}. It orders antiferromagnetically at
$T_N=9.2$K. The ordered local magnetic moment at zero field was
found to be equal to 0.15$\mu_B$ \cite{kenzelman-prb-2001}. The
easy axis of magnetization is aligned along the $c$ direction of
the orthorhombic crystal. This is confirmed by magnetization
measurements \cite{tsukada-prl-2001} and by neutron diffraction
results \cite{zheludev-prb-2002}. Instead of the expected single
spin-flop transition, two spin-flop transitions were observed on
the magnetization curves when the field was applied along the easy
axis $\myvec{H}||c$ \cite{tsukada-prl-2001} at $\mu_0 H_{c1}=2.0$T
and $\mu_0 H_{c2}=4.9$T. Later, ultrasonic studies
\cite{poirier-prb-2002} have revealed another phase transition
when the field was applied perpendicularly to the easy axis
$\myvec{H}||b$, with a critical field value of $\mu_0
H_{c3}=7.8$T. Finally, a spin-reorientation transition in the
third orientation $\myvec{H}||a$ was observed by
electron-spin-resonance \cite{glazkov-prb-2005} with a critical
field of $\mu_0 H_{c4}=11$T. The exact reason for these numerous
phase transitions is not completely clear yet. It was suggested
that a strong reduction of the ordered moment could increase the
importance of the anisotropy of the transverse susceptibility
\cite{glazkov-prb-2005}.

Despite these unusual properties, a detailed characterization of
the magnetic phase diagram of this compound remains incomplete.
The phase diagram in the $\myvec{H}||c$ orientation was determined
by magnetization measurements \cite{tsukada-prl-2001} and
ultrasonic studies \cite{poirier-prb-2002}. For the $\myvec{H}||b$
orientation, only ultrasonic data are available
\cite{poirier-prb-2002}. Finally, the phase transition in the
$\myvec{H}||a$ orientation was reported only at a temperature of
1.5K in electron-spin resonance experiments
\cite{glazkov-prb-2005}.  No magnetization studies of the phase
transitions for $\myvec{H}||a,b$ were reported to date.

In the present paper we fill in these gaps and report results of
magnetization and specific heat studies of the phase diagram of
\bacusio{}.

\begin{figure}
  \centering
  \epsfig{file=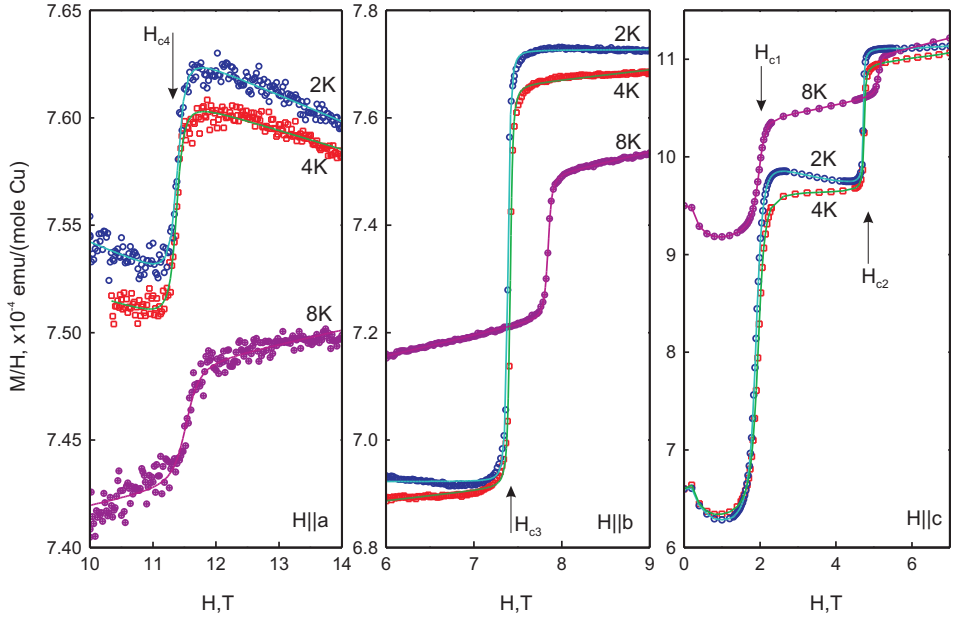, width=\figwidth, clip=} \caption{Field
  dependences of the M/H ratio in different orientations. The curves are
  guides for the eye.}\label{fig:m(h)}
\end{figure}

\begin{figure}
  \centering
  \epsfig{file=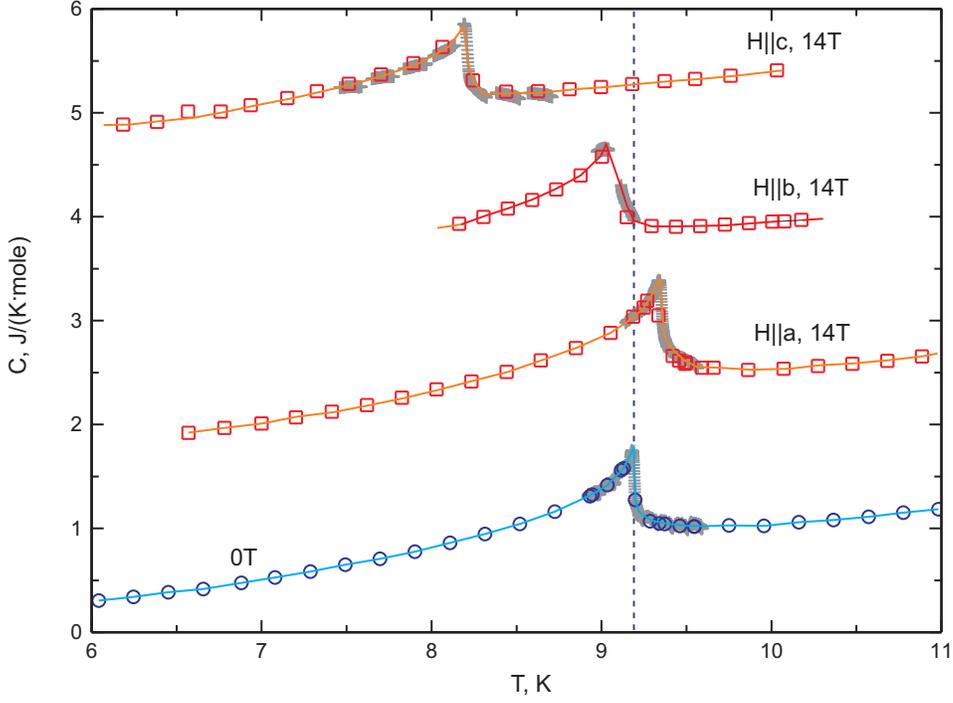, width=\figwidth, clip=}
  \caption{Change of the N\'{e}el temperature with field in
  different orientations. The specific heat curves measured at 14
  T for $\myvec{H}||a,b$ and $\myvec{H}||c$ are shifted upwards by
  1.5, 3.0 and 4.5 J/(K$\cdot$mole), respectively. Crosses:
  specific heat values obtained from slope analysis of thermal
  relaxation curves. Circles and squares: specific heat values
  obtained from best fits in the two-$\tau$ model. Vertical dashed
  line: position of the phase transition in zero field,
  $T_N=9.19$K. Solid lines: guides for the
  eye.}\label{fig:14T_c(t)}
\end{figure}

\begin{figure}
  \centering
  \epsfig{file=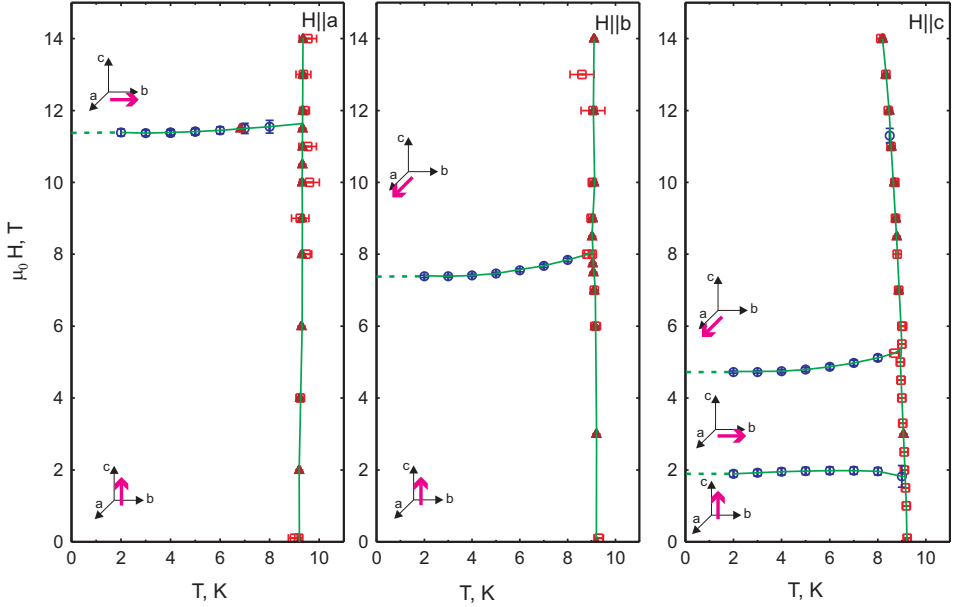, width=\figwidth, clip=}
  \caption{Complete set of phase diagrams for \bacusio{}. Circles:
  data from magnetization field scans. Squares: data from
  magnetization temperature scans. Triangles: data from specific
  heat measurements. Solid and dashed lines marking phase
  boundaries: guides for the eye. Orientation of the antiferromagnetic
  order parameter in all phases is shown according to Ref.
  \cite{glazkov-prb-2005}}\label{fig:t-h}
\end{figure}

\section{ Sample preparation and experimental details.}
We used single crystalline samples of \bacusio{} taken from the
same batch as those used in Ref.\cite{glazkov-prb-2005}. All
measurements were performed on the same sample of approximate size
$2\times3\times0.5$ mm$^3$. Sample orientation was checked by
X-ray diffraction on a Bruker APEX-II diffractometer. The
misalignment of the mounted sample is estimated to be within
5$^\circ$.

Magnetization was measured by a Quantum Design MPMS-XL
SQUID-magnetometer and a Quantum Design PPMS system equipped with
a vibrating sample magnetometer.

Specific heat was measured using Quantum Design PPMS system. It
was measured by applying a controlled heat pulse to the platform
with the sample, which is connected to the cryostat with a stable
heat-link. The sample temperature relaxation curve $T(t)$ was then
fitted by a two-$\tau$ model, yielding the value of the specific
heat. To improve the precision of the transition temperature
determination, a slope analysis technique was additionally used.
This technique makes use of the thermal balance equation
$P=C{\frac{dT_{sample}}{dt}}+K(T_{sample}-T_{cryo})$ (where $C$ is
the total specific heat of the sample and the platform, $P$ ---
the heater power, $K$ --- the heat link thermal conductivity,
$T_{sample}$ and $T_{cryo}$ --- being the temperatures of the
sample and of the cryostat, respectively).

\section{Experimental results and discussion.}

    Examples of magnetization  and specific heat curves are shown in Figures \ref{fig:m(h)} and
\ref{fig:14T_c(t)}. Both demonstrate clear anomalies at the phase
transitions. This allows to build phase diagrams in all
orientations of the applied magnetic field, as shown in Figure
\ref{fig:t-h}.

The zero-field transition temperature is
$T_N|_{H=0T}=9.19\pm0.01$K. The transition temperature is field
dependent, without strong anomalies at the crossings of the phase
transition lines. For $\myvec{H}||a$, the transition temperature
slightly increases with field, while for the other two principal
orientations, it decreases with field. The transition temperatures
at a field of 14T are: for $\myvec{H}||a$
$T_N|_{H=14T}=9.35\pm0.01$K, for $\myvec{H}||b$
$T_N|_{H=14T}=9.09\pm0.02$K, for $\myvec{H}||c$
$T_N|_{H=14T}=8.21\pm0.02$K.

The low-temperature values of the critical fields, as determined
by our magnetization measurements, are: for $\myvec{H}||c$
$H_{c1}=(1.89\pm0.10)$T, $H_{c2}=(4.72\pm0.03)$T; for
$\myvec{H}||b$ $H_{c3}=(7.39\pm0.03)$T; and for $\myvec{H}||a$
$H_{c4}=(11.40\pm0.08)$T. These values are in good agreement with
published data
\cite{tsukada-prl-2001,poirier-prb-2002,glazkov-prb-2005}. The
critical field values are weakly temperature dependent: the
corresponding phase transition lines are almost horizontal.

The measured magnetic susceptibilities of the higher-field phases
are always higher than those of lower-field phases (see Figure
\ref{fig:m(h)}). This confirms the identification of the
field-induced phase transitions as spin-reorientation transitions,
caused by the competition between the  order parameter anisotropy
and the Zeeman energy. Susceptibility jumps
$\Delta\chi=\Delta(M/H)$ at these transitions, as measured at 2K,
are: for $\myvec{H}||c$ $(3.54\pm0.05)\cdot 10^{-4}$emu/(mole~Cu)
and $(1.35\pm0.05)\cdot10^{-4}$emu/(mole~Cu), for the first and
second spin-flops, respectively; for $\myvec{H}||b$
$(0.813\pm0.020)\cdot10^{-4}$emu/(mole~Cu); and for $\myvec{H}||a$
$(0.085\pm0.020)\cdot10^{-4}$emu/(mole~Cu).

A model that describes all low-temperature phase transitions and
antiferromagnetic resonance frequency-field dependences was
proposed in Ref.\cite{glazkov-prb-2005}. This model suggests that
anisotropic contributions to the transverse susceptibility are
unusually large in \bacusio{}, probably due to the strong
reduction of the ordered magnetic moment. The low-energy dynamics
of the antiferromagnetic order parameter is then described by the
potential energy:

\begin{eqnarray}
U&=&-\frac{1}{2}\vectorprod{l}{H}^2+a_1l_x^2+a_2l_y^2+\xi_1\scalarprod{H}{l}H_xl_x
+\nonumber\\
&&+\xi_2\scalarprod{H}{l}H_yl_y
-(\xi_1+\xi_2)\scalarprod{H}{l}H_zl_z-\nonumber\\
&&-B_1H_x^2(l_y^2-l_z^2)-B_2H_y^2(l_x^2-l_z^2)-B_3H_z^2(l_x^2-l_y^2)+\nonumber\\
&&+C_1H_yH_zl_yl_z+C_2H_xH_zl_xl_z+C_3H_xH_yl_xl_y\label{eqn:energy}
\end{eqnarray}

\noindent Here, Cartesian coordinates are chosen as $x||a$, $y||b$
and $z||c$ and $\myvec{l}$ is the antiferromagnetic order
parameter. The exchange part of the transverse susceptibility is
set to unity for the sake of convenience. The phase transitions
are described in terms of this potential energy as rotations of
the order parameter $\myvec{l}$. In the low-field phases, the
order parameter is aligned along the easy axis, $\myvec{l}||z$.
When the field is applied along  the easy axis $z$, it rotates  at
$H_{c1}$ towards the second easy axis, $\myvec{l}||y$, and at
$H_{c2}$, towards the hard axis $\myvec{l}||x$. When  the field is
applied along $y$ or $x$, the order parameter rotates towards the
$x$ or the $y$ axis at $H_{c3}$ and $H_{c4}$, respectively. These
orientations of the order parameter are shown schematically in
Figure \ref{fig:t-h}.

The parameter values, all determined from best fits of the
electron spin resonance data \cite{glazkov-prb-2005}, are:
$\gamma=2.82$GHz/kOe, $a_1=400$kOe$^2$, $a_2=118$kOe$^2$,
$B_1$=0.0047, $B_2=0.0370$, $B_3=0.0614$, $\xi_1=0.135$,
$\xi_2=-0.03$. The $C_i$ constants  could be ignored in the
principal orientations of the magnetic field.

Our magnetization measurements allow an independent check of this
model, since susceptibility jumps at the phase transitions are
related to certain parameters of the potential (\ref{eqn:energy}).
Predicted susceptibility jumps at spin-reorientation transitions
are:
\begin{eqnarray}
\Delta\chi_1&=&1-2B_3-2(\xi_1+\xi_2)\\
\Delta\chi_2&=&4B_3\\
\Delta\chi_3&=&4B_2\\
\Delta\chi_4&=&4B_1
\end{eqnarray}

\noindent Here, $\Delta\chi_i$ are susceptibility jumps at the
corresponding critical field $H_{c~i}$. In order to exclude the
scaling factor involved in the choice of energy units in
Eqn.\ref{eqn:energy} they can be normalized to $\Delta\chi_1$.

A comparison of the calculated and measured values is given below:

\begin{tabular}{|c|c|c|}
    \hline
    &measured&calculated\\
    \hline
    $\Delta\chi_2/\Delta\chi_1$&$0.381\pm0.015$&0.365\\
    $\Delta\chi_3/\Delta\chi_1$&$0.230\pm0.007$&0.220\\
    $\Delta\chi_4/\Delta\chi_1$&$0.024\pm0.006$&0.028\\
    \hline
\end{tabular}

\noindent The correspondence between experimental and model values
is close to perfect. Thus, the magnetization measurements are
fully compatible with the proposed form of potential energy. This
supports the model of phase transitions proposed in
Ref.\cite{glazkov-prb-2005} and points to the non-trivial effects
of spin-reduction on spin-reorientation transitions in
low-dimensional antiferromagnets.

\ack The work was supported by RFBR Grant No.09-02-00736.

\end{document}